# A LINEAR ALGORITHM FOR THE GRUNDY NUMBER OF A TREE

Ali Mansouri[1] and Mohamed Salim Bouhlel [2]

[1]Department of Electronic Technologies of Information and Telecommunications Sfax, Tunisia
mehermansouri@yahoo.fr

[2]Department of Electronic Technologies of Information and Telecommunications Sfax, Tunisia
medsalim.bouhlel@enis.rnu.tn

## ABSTRACT

*A coloring of a graph G = (V ,E) is a partition $\{V_1, V_2, \ldots, V_k\}$ of V into independent sets or color classes. A vertex $v \in V_i$ is a Grundy vertex if it is adjacent to at least one vertex in each color class Vj for every j <i. A coloring is a Grundy coloring if every color class contains at least one Grundy vertex, and the Grundy number of a graph is the maximum number of colors in a Grundy coloring. We derive a natural upper bound on this parameter and show that graphs with sufficiently large girth achieve equality in the bound. In particular, this gives a linear-time algorithm to determine the Grundy number of a tree.*

## 1. INTRODUCTION

In this paper, we consider the calculation of the Grundy number of a graph with large girth. The concept of Grundy coloring is a weak form of minimal coloring.

Let G = (V ,E) be a graph. A set of vertices $S \subseteq V$ is called independent if no two vertices in S are adjacent. A (proper) k-coloring of graph G is a partition $\{V_1, V_2, \ldots, V_k\}$ of V into k independent sets, called color classes. In the coloring, a vertex $v \in V_i$ is called a Grundy vertex if v is adjacent to at least one vertex in color class $V_j$, for every j <i. (Every vertex in $V_1$ is vacuously a Grundy vertex.) Thus the Grundy (coloring) number of a graph is defined as the maximum number of colors in a coloring where every vertex is a Grundy vertex. This parameter is studied in [1–3,5–8,11] inter alia.

Recently, Erdös et al. [4] studied the concept of a partial Grundy coloring. This is a coloring $(V_1, V_2, \ldots, V_k)$ such that every color class Vi, 1<i<k, contains at least one Grundy vertex. The Grundy number $\partial \Gamma (G)$ is the maximum k such that G has a k-Grundy coloring [4]. Erdös et al. related Grundy colorings to other graph properties such as parsimonious proper coloring number [10] and the maximum degree.

In this paper, we observe a natural upper bound on $\partial \Gamma (G)$ and show that trees and graphs with sufficiently large girth achieve equality in the bound. In particular, this gives a linear time algorithm to determine the Grundy number of a tree. As expected, the associated decision problem is NP-complete

## 2. AN UPPER BOUND

In this section, we define a natural upper bound on the Grundy number in terms of what we call a "feasible Grundy sequence". We show that a greedy algorithm can compute this bound on all graphs.

**Definition**. A sequence S of r distinct vertices $(g_1, \ldots, g_r)$ of a graph G is a feasible Grundy sequence if for $1 < i < r$ the degree of $g_i$ in $G - \{g_{i+1}, \ldots, g_r\}$ is at least $i - 1$. The stair factor $\ell(G)$ of G is the maximum cardinality of a feasible Grundy sequence.

For example in the graph in Fig. 1, the sequence (b, c, d, e) is a feasible Grundy sequence.

Note that the stair factor is clearly no greater than the maximum degree plus 1. If we have a Grundy coloring and write down a Grundy vertex from each color class from $V_k$ down to $V_1$, we obtain a feasible Grundy sequence. Thus it follows that:

$\partial \Gamma(G) < \ell(G) < \Delta(G) + 1$.

Surprisingly perhaps, $\ell(G)$ can be calculated in linear-time with a simple greedy algorithm. (This has echoes in the calculation of the so-called coloring number of a graph; see [9]). We introduce the following lemma.

**Lemma 2.1**. Let w be a vertex of graph G. Then $\ell(G - w) > \ell(G) - 1$.

**Proof**. Consider a maximum feasible Grundy sequence S of graph G: say $(g_1, \ldots, g_?)$. If $w \in S$, then $S - w$ is a feasible Grundy sequence in graph $G - w$. If $w \notin S$, then $S - g_1$ is a feasible Grundy set in $G - w$. Hence the result.

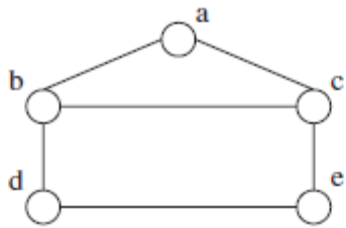

Figure 1. An example of feasible Grundy sequence.

We next introduce a vertex-decomposition list.

**Definition**. A vertex-decomposition list $D = (v_1, \ldots, v_n)$ of a graph is computed by repeatedly removing a vertex $v_i$ of maximum degree from the graph $G_i = G - \{v_1, \ldots, v_{i-1}\}$. (Note that $v_1$ is a maximum degree vertex in G.) The degree of the vertex $v_i$ at removal we call its residue degree, and denote the sequence of residue degrees by $(d_1, \ldots, d_n)$.

**Theorem 1**. Let D be any vertex-decomposition list of a graph G with residue degrees $(d_1, \ldots, d_n)$. Then

$\ell(G) = \min d_i + i$.

**Proof**. Let k be the value $\min_i d_i + i$. It is clear that $(v_1, \ldots, v_k)$ is a feasible Grundy sequence. Hence $\ell(G) > k$.

We prove that ל (G) < k by induction on the order n. If n=1, then the graph is an isolated vertex, and so ל (G) = k = 1. In general, let G'= G − $v_1$ and let k' be the value produced by the algorithm on G' using D' = D − $v_1$. That is, k' = $\min_{i>2} d_i + (i − 1)$.

There are two cases.

**Case 1**: k'< k. Then ל (G) < k' by the inductive hypothesis, and so ל (G) < ל (G') + 1 = k' + 1< k by Lemma 2.1.

**Case 2**: k' = k. Then the only way that $\min_{i>2} d_i + i − 1 = \min_{i>1} d_i + i = k$ is that $d_1 = k − 1$. That is, Δ (G) = k − 1. But clearly ל (G) < Δ (G) + 1 in all graphs.

## 2.1. Complexity analysis

We show that the parameter ל (G) can be computed in O (m) time, where m is the number of edges in the graph. We assume the graph is given by its adjacency list representation.

We proceed as follows. We use bucket sort to distribute the n vertices into an array of n buckets corresponding to residue degree 0 through n − 1. While calculating the decomposition sequence, we find the vertex $v_i$ in the last non-empty bucket and remove it. Then for each entry u in $v_i$ 's adjacency list, we move u one bucket down.

To allow each bucket move to take O (1) time, we store each bucket as a doubly linked list; and also maintain for each vertex vi a pointer that tracks the location of its entry in the bucket-array. During a run of the algorithm, the total number of moves is at most the sum of the degrees and is hence O (m). Therefore the algorithm has complexity O (m).

## 3. ACHIEVING GRUNDY COLORING

In this section we show that, in graphs with sufficiently large girth, any feasible Grundy sequence is always "almost realizable" in a Grundy coloring, and present a linear time algorithm for achieving this. This implies that on such graphs the partial Grundy number is equal to the upper bound. We prove the following theorem.

**Theorem 2**.

Let G be a graph with girth larger than 8 and let S=($g_1$, . . . , $g_k$) be a feasible Grundy sequence for G. Then there exists a partial Grundy coloring ($V_1$, . . . , $V_l$) with l > k such that for each i >2, $g_i$ is in Vi and is a Grundy vertex.

Note that the restriction i>2 is necessary: consider for example a path on 4 vertices labeled $g_1$–$g_3$–v–$g_2$. In this case, the vertex labeled g1 is given color 2 following

**Theorem 2.**

We first give the following definition.

**Definition**. Consider a feasible Grundy sequence ($g_1$, . . . , $g_k$). Vertex $g_i$ is satisfied if we properly color some of its neighbors in such a way that $g_i$ is adjacent to colors {1, . . . , i−1},but i is not among the colors now appearing on its neighbors. (In other words, $g_i$ becomes a Grundy vertex for color class $V_i$)

The basic approach in proving Theorem 2 is as follows:

We color the graph in k −1 rounds. In Round i (i running from 2 up to k), we color the vertex gi with color i and then satisfy it. At the end, we color all the uncolored vertices (if any) one at a time by giving each the smallest color not used in its neighborhood.

To satisfy $g_i$, we must color i−1 of its neighbors with distinct colors. So we must ensure that each round of coloring does not create problems for later rounds: specifically it must ensure that already-colored neighbors have distinct colors. For this we introduce the concept of a restriction.

**Definition**. Let w be an uncolored neighbor of $g_i$. Then a restricted color for w, is any color at a neighbor of w, and any color at distance 2 from w, where the common neighbor of w and the colored vertex is one of $g_{i+1}, \ldots, g_k$.

Note that at the start of round i, every colored vertex is either one of $g_2, \ldots, g_{i-1}$, or has been colored to satisfy one of these vertices (its coloring was caused by that Grundy vertex). Thus there are 4 cases of restricted colors,. (Note that in case (1), w was not used to satisfy its adjacent Grundy vertex.) The key is that the girth limits the number of restrictions. Note that when restrictions are not violated, no two vertices in Ni (see below) are already colored with the same color at round i.

**Lemma 3.1**. Let graph G have girth larger than 8. Let Ni be any set of i − 1 neighbors of gi excluding $g_{i+1}, \ldots, g_k$. If at the start of round i no two vertices in $N_i$ are already colored with the same color, then we can complete the coloring of Ni such that every vertex in $N_i$ has a distinct color in the range $\{1, \ldots, i − 1\}$ and the color restrictions are not violated.

**Proof.** Assume that no restrictions have been violated up to the start of round i. Before round i, only colors 1 up to i − 1 have been used. Consider a vertex $g_j$ for 2< j <i. Then, by the girth constraint, either $g_j$ caused one colored neighbor of $g_i$ and no restriction on the uncolored vertices of Ni, or $g_j$ caused no colored neighbor and at most one restriction.

Hence if b of the elements of $N_i$ are uncolored, then there are at most b−1 restrictions and b available colors. It follows that we can complete the coloring of Ni greedily—provided we color the vertices that have a restriction first; each time, we remove the color used from the available list at each vertex. Because the restrictions are not violated and available list is updated after each coloring, no two vertices in $N_i$ are colored with the same color while satisfying $g_i$. No restrictions have been violated in the execution of round i.

We observe that a set N always exists. So this completes the proof of Theorem 2. As a consequence, we obtain:

**Theorem 3**. For a graph G of girth larger than 8, ל(G) = ∂ Γ (G).

## 4. THE GRUNDY (COLORING) NUMBER

As expected, the Grundy number is intractable. Define the following decision problem:

GRUNDY COLORING:

Instance: Graph G, positive integer k

Question: Does G have a partial Grundy coloring $\prod = \{V_1, V_2, \ldots, V_k\}$ with at least k colors?

**Theorem 4**. GRUNDY COLORING is NP-complete, even for chordal graphs.

**Proof.** Clearly GRUNDY COLORING is in NP. The certificate is the coloring.

The transformation is from 3-COLOR (is a graph 3-colorable?). Given an arbitrary instance G of 3-COLOR, we create a graph G' and integer k, such that G has a (proper) 3-coloring if G' has a Grundy k-coloring for G'.

The graph G' is constructed as follows:

1. For each vertex $v_i \in V(G)$ create a single vertex labeled $v_i$ in G'. For each edge $e_j \in E(G)$ create a single vertex labeled $l_j$ in G'. Let R be the set of these vertices (both $v_i$ and $l_j$) in G'.

2. For each edge $e_j = v_a v_b \in E(G)$, create a vertex $s_j$ in G', and add edges from $s_j$ to $v_a, v_b$ and $l_j$ in G'. Then form all the $s_j$ vertices into a clique S in G'.

3. Finally, add one disjoint $K_3$ in G'.

An example of this construction is shown in Fig. 2.

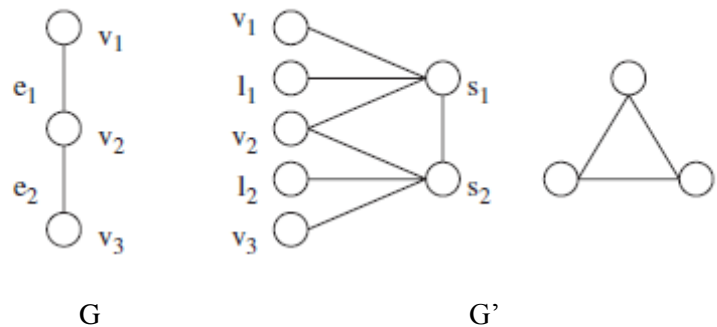

Figure 2. An example

Clearly, this construction is polynomial in the size of G. Let $k = |E| + 3$ (the maximum degree of G' plus 1).

First we show that if G has a 3-coloring g, then G' has a Grundy coloring f that uses k colors. We construct f as follows:

1. For $v_i \in V$ let $f(v_i) = g(v_i)$. For each $l_j$ in G' representing an edge $e_j = v_a v_b$ in G, let $f(l_j)$ be whichever of 1, 2 or 3 is not used for $v_a$ and $v_b$.

2. For $j = 1, \ldots, |E|$, let $f(s_j) = 3 + j$ (in any order).

3. Color the $K_3$ with 1, 2 and 3.

The assignment is a valid partial Grundy k-coloring. For, each color class $V_i$ is an independent set. Note that for $i > 4$, each color class $V_i$ has only one member, namely $s_{i-3}$.

Each $s_j$ vertex is adjacent to vertices colored 1, 2 and 3 in R ($v_a$, $v_b$, and $l_j$) and to vertices colored $4, \ldots, j-1$.

Now we show that if G' has a Grundy coloring f using k colors, then G has a proper 3-coloring. Indeed, we claim that the restriction of f to V is a proper 3-coloring of G. Note the following:

1. Suppose some vertex of R is Grundy in G'. Let w be such a vertex with the largest color c, c>1. Vertex w is only adjacent to vertices of S; hence S must contain vertices of colors 1 through c − 1. By the choice of w, S contains Grundy vertices for colors c + 1 through k. This implies that |S| > k − 1, a contradiction. Therefore, in graph G', no vertex of R is Grundy with color larger than 1. [12]

2. Thus S consists of the Grundy vertices for colors 4 up to k. Each of these vertices has three neighbors in R: these must be colored with colors 1, 2 and 3, and must have distinct colors.

3. Thus V is colored with only colors 1, 2 and 3. Furthermore, if $v_a$ and $v_b$ are adjacent in G, there is a vertex of S adjacent to both of them and so they receive different colors. That is, f restricted to V is a proper coloring of G.

A similar construction shows that the partial Grundy number is intractable even for bipartite graphs.

## 4. CONCLUSION

In this paper we show that the problem of determining the Grundy number $\partial \Gamma (G)$ of a graph G is NP-complete even for chordal graphs and bipartite graphs. We also show that $\partial \Gamma (G)$ for any G is bounded above by the stair factor ל(G), a parameter newly defined here.

For graphs with girth larger than 8, this bound is achieved. This result leads to a linear time algorithm to determine $\partial \Gamma (G)$ for graphs with large girth.

We also observe that $\partial \Gamma (G) <$ ל$(G) < \Delta (G) + 1$ for any graph G. This inequality chain raises some interesting questions. We know from Theorem 3 that for a graph G with girth larger than 8, $\partial \Gamma (G) =$ ל$(G)$.

## ACKNOWLEDGEMENTS

I think Mr. Zhengnan Shia for his collaboration.